# Revealing Atomic Site-dependent g-factor within a Single Magnetic Molecule via Extended Kondo Effect

Liwei Liu,[1†] Kai Yang,[1†] Yuhang Jiang,[1†] Boqun Song,[1†] Wende Xiao,[1] Shiru Song,[1] Shixuan Du,[1] Min Ouyang,[2] Werner A. Hofer,[3] Antonio H. Castro Neto[4], and Hong-Jun Gao[1*]

[1] *Institute of Physics, Chinese Academy of Sciences, P.O. Box 603, Beijing 100190, China*

[2] *Department of Physics and Center for Nanophysics and Advanced Materials, University of Maryland, College Park, MD 20742, USA*

[3] *School of Chemistry, Newcastle University, Newcastle, UK*

[4] *Graphene Research Centre, Department of Physics, National University of Singapore, 117542, Singapore*

Site-dependent g-factor of a single magnetic molecule, with intramolecular resolution, is demonstrated for the first time by low-temperature, high-magnetic-field scanning tunneling microscopy of dehydrogenated Mn-phthalocyanine molecules on Au(111). This is achieved by exploring the magnetic-field dependence of the extended Kondo effect at different atomic sites of the molecule. Importantly, an inhomogeneous distribution of the g-factor inside a single molecule was revealed. Our results open up a new route to access local spin properties within a single molecule.

The magnetic properties of a nanostructure play a pivotal role in the design of miniaturized spintronic devices [1]. The Kondo resonance, due to many-body spin-flip scattering between a local magnetic moment and the conduction electrons of a host metal, has been increasingly popular in the scientific community in recent years [2, 3]. One signature of a Kondo resonance is its splitting at the presence of an external magnetic field, doubling the usual Zeeman splitting [4]. This field-induced Kondo splitting can be used to characterize the g-factor of systems like quantum dots in transport or scanning tunneling microscopy (STM) experiments [5-9]. Importantly, the effective g-factor (or its deviation from that of free electrons) depends not only on magnetic anisotropy but also on the local spin-orbit coupling [10, 11]. For example, spin-orbit interaction has been found to cause a contribution of orbital angular momentum to appear in the ground state [11]. Thus, the ability to probe the fine structure of the g-factor should allow us to understand the internal spin properties of a complex molecule, if the Kondo splitting can be locally resolved inside a molecule.

Here, we report a systematic investigation of the Kondo effect in a series of dehydrogenated (DH-) Mn-phthalocyanine (MnPc) molecules, which are modified through detachment of hydrogen atoms by STM. We observe that the Kondo effect is extended in space beyond the central Mn ion, and onto the non-magnetic constituent atoms of the molecule. This extended Kondo effect can be explained by spin polarization induced by symmetry breaking of the molecular framework, as confirmed by density functional theory (DFT) calculations. Measuring the evolution of the Kondo splitting with applied magnetic fields at different atomic sites, we find a spatial variation of the g-factor within a single molecule for the first time. The existence of atomic site-dependent g-factors can be attributed to different spin-orbit coupling of molecular orbitals within the molecule. As both, molecular orbitals and their associated g-factor are relevant for the chemical environments, our results provide a new route to explore the internal electronic and spin structure of complex molecules, hard to achieve otherwise.

The MnPc molecules were deposited on an Au(111) substrate at room temperature, and the sample was subsequently cooled down to 0.4 K for all STM measurements in this report. The molecular structure of MnPc is shown in the left panel of Fig. 1(a). On a terrace, each MnPc molecule appears as a four-lobed cross with a central protrusion (see right panel of Fig. 1(a), and Fig. S1(a) in Ref. [12]), which is consistent with molecular $D_{4h}$ symmetry and indicates a flat-lying adsorption configuration. Subsequently, we have detached single hydrogen atoms by STM manipulation, to sequentially alter the molecular structure and the electronic properties of the molecule [13-15]. This can be achieved by placing the STM tip above a specific molecular lobe (as indicated by red dotted circles in the left panel of Fig. 1(a)) and then applying a voltage pulse of 3.6 V for one second with the feedback loop disengaged. Figure 1(b) shows consecutive topographic images of dehydrogenated MnPc molecules (see also Fig. S1(b) in Ref. [12]). We denote the as-modified molecules as -2H-MnPc, -4H-MnPc, -6H-MnPc, -8H-MnPc, for the removal of two, four, six and eight hydrogen atoms, respectively.

Together with changes in the topography, the *dI/dV* spectra acquired at the central Mn ion of dehydrogenated molecules show an evolution of the local density of states (LDOS) near the Fermi level [Fig. 1(c)], from a step-like feature in the complete MnPc molecule (bottom curve in Fig. 1(c)) to a sharp peak for -6H-MnPc molecule and a double-peak structure for the -8H-MnPc molecule (top curve in Fig. 1(c)). The new spectral features of the dehydrogenated molecules can be attributed to the characteristics of the Kondo resonance, due to spin-flip scattering between the local magnetic impurity, the Mn ion, and the conduction electrons of the substrate [16-21]. The presence of a zero-field split (ZFS) of the Kondo resonance for the -8H-MnPc is due to the magnetic anisotropy experienced by the central Mn ion.

One of the hallmarks of a Kondo resonance is its splitting in the presence of an external magnetic field B [4-9]

$$\Delta E = 2g\mu_B B, \qquad (1)$$

where $\mu_B$ is the Bohr magneton. In the spectroscopy experiments we observe the split in the *dI/dV* spectra under an applied magnetic field. The results are summarized in Figs. S2 and S3 in Ref. [12], confirming the existence of the Kondo resonance.

To quantitatively analyze the evolution of the Kondo resonance as the molecule is dehydrogenated, we have fitted the *dI/dV* spectra by taking two Fano functions into account [20]:

$$\frac{dI}{dV}(V) = A \cdot F(V) + B \cdot F\left(V - \frac{\Delta}{e}\right) + C, \qquad (2)$$

$$F(V) = \frac{(\varepsilon + q)^2}{1 + \varepsilon^2}, \qquad (3)$$

where $\varepsilon = (eV - \varepsilon_0)/\Gamma$ is the normalized energy ($\varepsilon_0$ is the position of the resonance. $\Gamma$ is the half-width at half-maximum of the Kondo peak, which is related to Kondo temperature $T_K$ as $\Gamma \sim k_B T_K$ at low temperatures), $\Delta$ is the Zeeman splitting energy of the Kondo peak, and $q$ is the Fano asymmetry factor given by $q \propto t_2/t_1$ ($t_1$ and $t_2$ are the matrix elements for electron tunneling into the continuum of the bulk states and the discrete Kondo resonance, respectively [22]). The use of two Fano functions is necessary in order to describe the spectra in the presence of ZFS or Zeeman splitting.

For the intact MnPc molecule an asymmetric Kondo resonance is detected,

yielding a negative *q* factor of -1.16 ± 0.1 and $T_K$ of 36 ± 1 K [Fig. 1(d)]. The Kondo temperature $T_K$ decreases from 36 K to 16 K with the increase of the number of removed peripheral H atoms. Because our sample temperature of 0.4 K for STM measurements is well below $T_K$ in all cases, we can conclude that the observed Kondo scattering occurs in the strong coupling regime [23]. The variation of the shape factor *q* with the number of detached H atoms agrees well with the variation of the Kondo temperature $T_K$.

To understand the observed variations of both, $T_K$ and *q* with the number of removed H atoms, we have performed DFT calculations on a series of dehydrogenated MnPc molecules. The DFT calculations suggest that the distance between the Mn ion and the substrate increases logarithmically from 3.2 Å for the intact MnPc to about 5 Å for the dehydrogenated MnPc [Fig. 1(e)], while the magnetic moment of the Mn ion is almost unchanged in the process of dehydrogenation. As the coupling strength between the Mn ions and the substrate atoms is governed by the distance between them, a larger distance results in a weaker interaction, and hence a reduced Kondo temperature according to $k_B T_K \propto e^{-1/\rho J}$ ($\rho$ is the density of states of the host and *J* is the spin-electron coupling) [24]. Meanwhile, increasing the distance between the molecule and substrate atoms decreases the probability of electrons tunneling from the STM tip into bulk continuum states, yielding an increase of the absolute value of *q*. Consequently, the distance evolution between molecule and substrate due to dehydrogenation of the molecule explains qualitatively the variations of $T_K$ and *q* with the number of removed H atoms, suggesting that the distance plays a crucial role in the variation of the Kondo resonance.

In order to evaluate the spatial distribution of the molecular Kondo resonance, we have performed *dI/dV* mapping over both, the intact and the dehydrogenated MnPc molecules near the Fermi level. For a Kondo system, the intensity of the *dI/dV* map at zero bias signifies the magnitude of the Kondo resonance at zero magnetic field. The *dI/dV* image of an intact MnPc shows a single and bright spot at the molecular center (see Fig. S4(a) in Ref. [12]), indicating that the Kondo resonance is fully localized at the Mn ion of the molecule. Our observation of a highly localized Kondo effect in intact MnPc molecules is consistent with previous work [18, 19]. A similarly localized Kondo feature is also observed for the -2H and -8H-MnPc molecules (see Figs. S4(b) and S4(c) in Ref. [12]).

By contrast, the *dI/dV* maps of both, the -4H-MnPc and the 6H-MnPc molecules show drastic variations from the intact molecules [Figs. 2(a) and 2(b)]. For example, two additional spots appear above the remaining lobes of the -4H-MnPc molecule. Figure 2(c) highlights two typical *dI/dV* spectra acquired at the lobe positions, which are manifestly due to Kondo resonances. Intuitively, the appearance of an *extended* Kondo effect suggests that a net spin polarization exists at the lobes. To check the validity of this assessment, we have performed DFT calculations for the -4H-MnPc and -6H-MnPc molecules. The projected spin density of states (PDOS) for the -4H-MnPc molecule clearly shows spin polarization not only at the position of the Mn ion but also at the benzene rings of the remaining two lobes [Fig. 2(d)]. DFT simulations for the -6H-MnPc molecule yield a similar picture, supporting the existence of an extended Kondo resonance [Fig. 2(e)].

Generally, the occurrence of local magnetic properties on otherwise non-magnetic atoms must be due to a change of the local potential environment. One possible origin is the symmetry breaking of the molecular framework and its effect on local potentials. DFT calculations show, for example, that the dehydrogenated lobes of -6H-MnPc molecule adopt a bending configuration with respect to the Au substrate [Fig. 3(a)], in contrast to the symmetric planar adsorption of an intact MnPc molecule [Fig. 1(a)]. This leads to a shift of the energy levels of the spin up and spin down electrons, as revealed by density of states projected onto the $p_x$, $p_y$ and $p_z$ orbitals of the C and N atoms (see Figs. 3(b), 3(c) and also S6 in Ref. [12]), resulting in a net spin polarization. Alternatively, the spin polarization may also be induced by charge transfer from the Au substrate [25]. Here, this mechanism can be excluded by calculating the charge density induced by molecular bonding to the substrate: it shows no net charge transfer from the substrate to the molecule [Fig. 3(a)] [26].

The existence of an extended Kondo resonance in a complex molecule offers an opportunity to evaluate the spatial distribution of the g-factors by exploring their magnetic field dependence. Here, we use the -6H-MnPc molecule as an example, and investigate in detail its electronic properties during a variation of the external magnetic field. When an external magnetic field is applied, all peaks in the *dI/dV* spectra of -6H-MnPc molecule split [Fig. 4(b)-(f)]. This supports our assignment of the spatially extended spectral feature to a Kondo effect, and excludes other possibilities due to molecular orbitals [28, 29]. The measured energy splitting follows a linear dependence with the applied magnetic field, which can be used to determine the corresponding g-factors (see Fig. S5 in Ref. [12]). Importantly, the g-factor exhibits a spatial variation over the molecule, and it is dependent on the constituent atomic species. The g-factors of the central Mn ion and the C atoms of the lobe are found to be 3.09 ± 0.56 and 1.73 ± 0.23, respectively. Similar field-dependent measurements performed on different nitrogen atoms surrounding the Mn ion (labeled N1, N2 and N3 for clarity) reveal g-factors of 1.97 ± 0.03, 1.82 ± 0.07 and 2.26 ± 0.04, respectively..

Quantitative model of distinct g-factors at different atomic sites within a molecule are currently under development. However, qualitatively our observation of a spatial variation of g-factors within the molecule can be attributed to the inhomogeneous local chemical and potential environment inside the molecule. The *d* orbitals of the molecule are mainly localized around the central Mn ion, with a net spin S = 3/2. Although the ligand field tends to quench the orbital angular momentum, the spin-orbit interaction mixes different states with the same orbital angular momentum, resulting in a ground state with non-zero orbital angular momentum. This results in an effective g-factor larger than 2 [10]. Due to symmetry breaking induced by the dehydrogenation as shown in Fig. 3, the non-magnetic macrocycle of the -6H-MnPc molecule also acquires a net spin, which is mainly contained within *p* orbitals [Figs. 2(e), 3(b) and 3(c)]. The net spin at N and C sites interacts mainly with the substrate electrons, and thus the g-factors are close to 2. Consequently, the spin moments residing on the *d* orbitals (at the Mn site), as well as on the *p* orbitals (at the N and C sites), couple to a different degree with the orbital angular momentum, resulting in a variation of the g-factor within the molecule. The validity of this - qualitative - model also implies a variation of the Kondo temperature at different atomic sites within the molecule. As Figure 4(h) shows, $T_K$ at different atomic sites is indeed different, supporting the validity of our qualitative model.

In summary, we have experimentally detected a spatial variation of the g-factor inside a single molecule for the first time. The results are obtained by measuring the Zeeman splitting of a Kondo resonance above specific atoms of the molecule. Qualitatively, the effect can be attributed to the symmetry-breaking of the molecular frame due to dehydrogenation of the molecule. Compared to other techniques like spin-polarized STM, our approach does not require a magnetic tip, and it directly enables to probe the g-factors within a single molecule. The spatial variation of the g-factor, which is due to the local potential environment, could in the future allow us to study local spin-orbit interactions, once the necessary theoretical tools have been developed.

**Figure Captions**

FIG. 1. Manipulation of the molecular Kondo effect by systematic dehydrogenation. (a) Molecular model (left) and STM images of intact MnPc molecules. Scanning bias: -0.2 V; tunneling current: 10 pA; scale bar, 1 nm. (b) STM images of dehydrogenated MnPc molecules. The pair of outer hydrogen atoms of each of the four lobes (as indicated by dotted circles in (a)) were removed step by step. Scanning bias: -0.2 V; tunneling current: 8.6 pA; scale bar, 1 nm. (c) Corresponding *dI/dV* spectra of molecules in (a) and (b). All spectra were acquired at the position of central Mn ions. Each successive plot is vertically shifted 0.2 nA/V for clarity. The red curves are Fano fittings of the Kondo resonance as described in main text. (d) Evolution of Kondo temperature ($T_K$) and *q* factor with the number of removed H atoms, derived from Fano fitting of *dI/dV* spectra. (e) Calculated distance between central Mn ion and Au substrate (the first layer) with the increase of number of H atoms removed. The red dashed line shows an exponential fit.

FIG. 2. Spin polarization properties of dehydrogenated MnPc molecules. (a) and (b) *dI/dV* mapings of -4H and -6H-MnPc molecules, respectively. *dI/dV* mappings were taken at 6 mV; scale bar, 1 nm. The dotted lines are molecular contours. (c) Typical *dI/dV* spectra of -4H and -6H-MnPc molecules. The spectra were acquired at the lobe indicated by the black cross in (a) and (b), respectively. The top curve is vertically shifted 0.2 nA/V for clarity. The black curves are Fano-fittings of the Kondo resonance. The Kondo line shape measured at the lobes differs slightly from the shape at the center, which could be due to different chemical environment of local spins within the molecule. (d) and (e) Calculated spin polarization of -4H-MnPc and -6H-

MnPc, respectively. The theoretical results are overlaid on top of experimental *dI/dV* mappings, showing the origin of extended Kondo effect. Spin polarization (P) is defined as P = (n↑ - n↓)/ (n↑ + n↓), where n↑ and n↓ are the LDOS of spin up and down electrons, respectively. The color scale of intensity of spin polarization is given below (d) and (e).

FIG. 3 Mechanism of observed extended Kondo effect. (a) Top and side views of the differential charge density induced by molecular bonding to the substrate. The excess (depletion) of charge is shown in yellow (green). (b) and (c) Calculated density of states of -6H-MnPc/Au(111) projected onto the orbitals of C and N1 atoms (The positions of the two atoms are labeled in Fig. 4(a)). Positive (negative) units refer to spin up (down) states. A Gaussian broadening of 0.05 eV has been employed in the calculation.

FIG. 4. Site-dependent g-factor of a -6H-MnPc molecule. (a) 3D stacking map showing topographic image, discrete g-factors (corresponding to Fig. 4g) and $T_K$ (corresponding to Fig. 4h) in different 2D layers. Topography of a -6H-MnPc molecule is overlaid with schematic molecular model. Scale bar, 0.5 nm. The contour plots (dashed lines) indicate the relative positions of the atoms with respect to the active Kondo areas. g-factors layer: Bar heights represent the g values with respect to the free electron g value (2). Green or red colors represent that the value is higher or lower than that of the free electron, respectively. (b) to (f), *dI/dV* spectra acquired at different atomic sites specified in (a). All *dI/dV* spectra were measured at the sample temperature of 0.4 K under a magnetic field of $B_z$ = 0～11 T. The zero-field splitting of the Kondo peak may be due to magnetic anisotropy [27]. The slightly asymmetric

shape of the split peaks arises from the asymmetric Fano-shaped Kondo resonance at zero magnetic field. Each successive plot is vertically shifted by 0.2 nA/V for clarity. (g) g-factors calculated from the Kondo splitting in (b) to (f) (see Fig. S5 in Ref. [12] for the linear fitting). (h) Kondo temperatures ($T_K$) extracted from Fano fitting of the spectra at different atomic sites of molecule under zero magnetic field. Note that the variations of $T_K$ and g-factors within the molecule do not necessarily need to follow the similar way.

**Acknowledgements**

Work at IOP was supported by grants from the National Science Foundation of China, National "973" projects of China, the Chinese Academy of Sciences, and NSCC. Work in Newcastle was supported by the EPSRC Car-Parinello consortium, grant No EP/F037783/1. M.O. acknowledges support from the ONR (award #: N000141110080 N000141410328) and NSF (award #: DMR 1307800). AHCN acknowledges DOE grant DE-FG02-08ER46512, ONR grant MURI N00014-09-1-1063, and the NRF-CRP award "Novel 2D materials with tailored properties: beyond graphene" (R-144-000-295-281).

# Figures

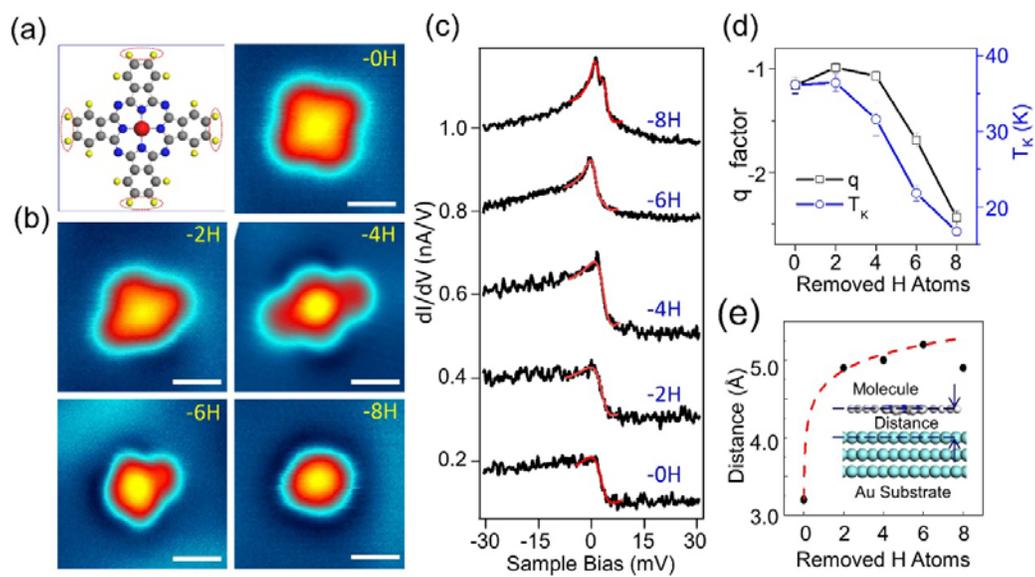

**FIG. 1**

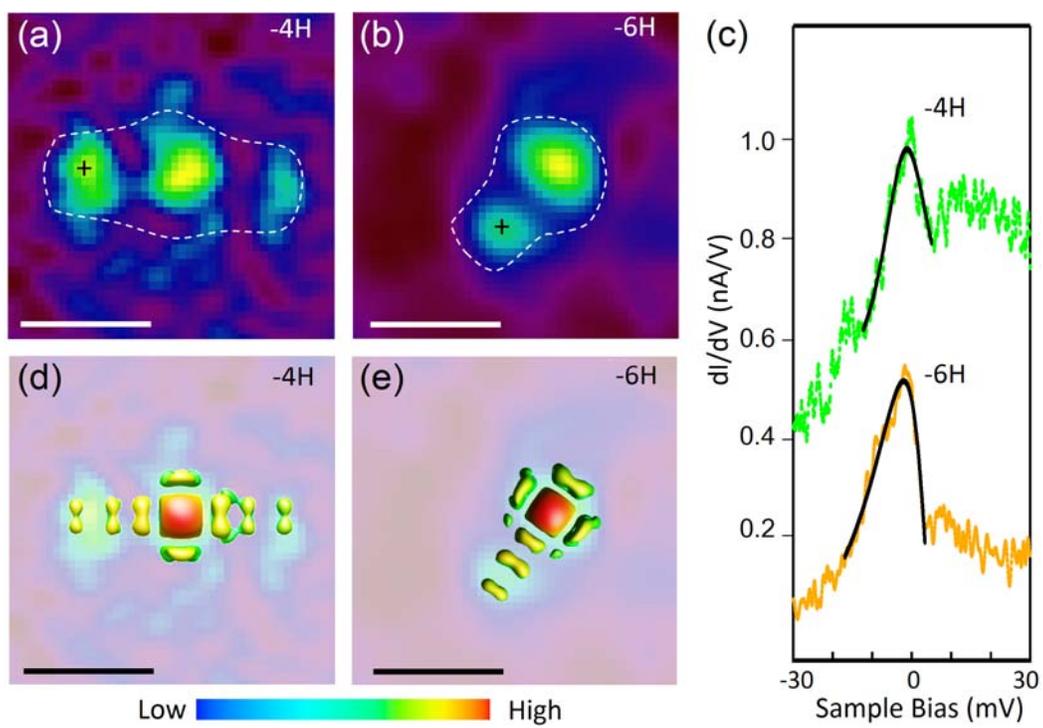

**FIG. 2**

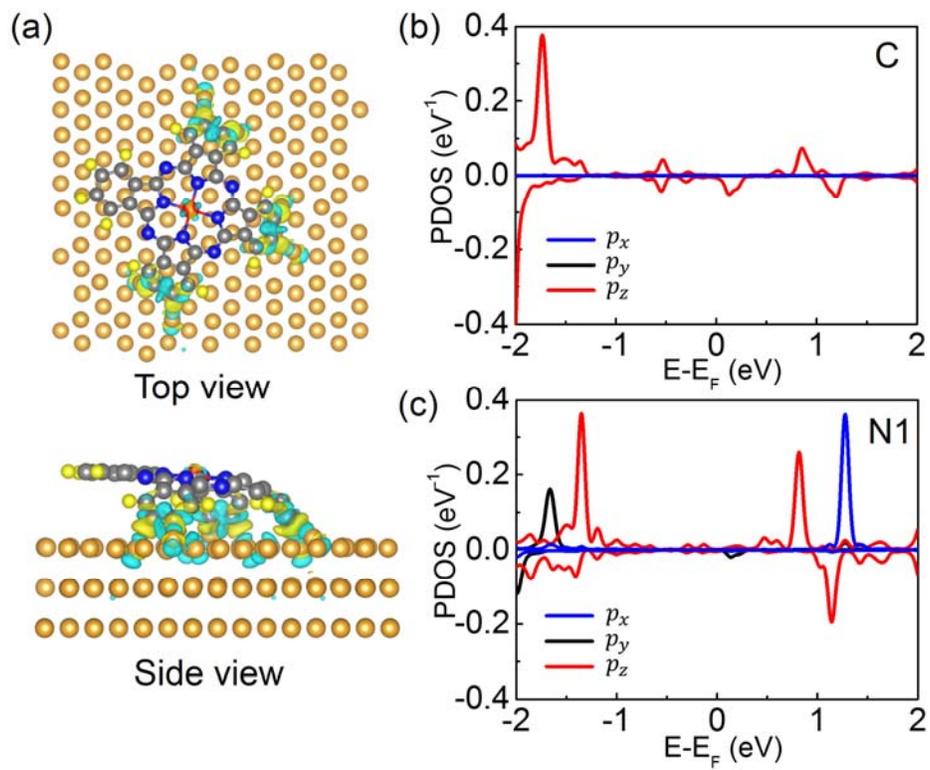

**FIG. 3**

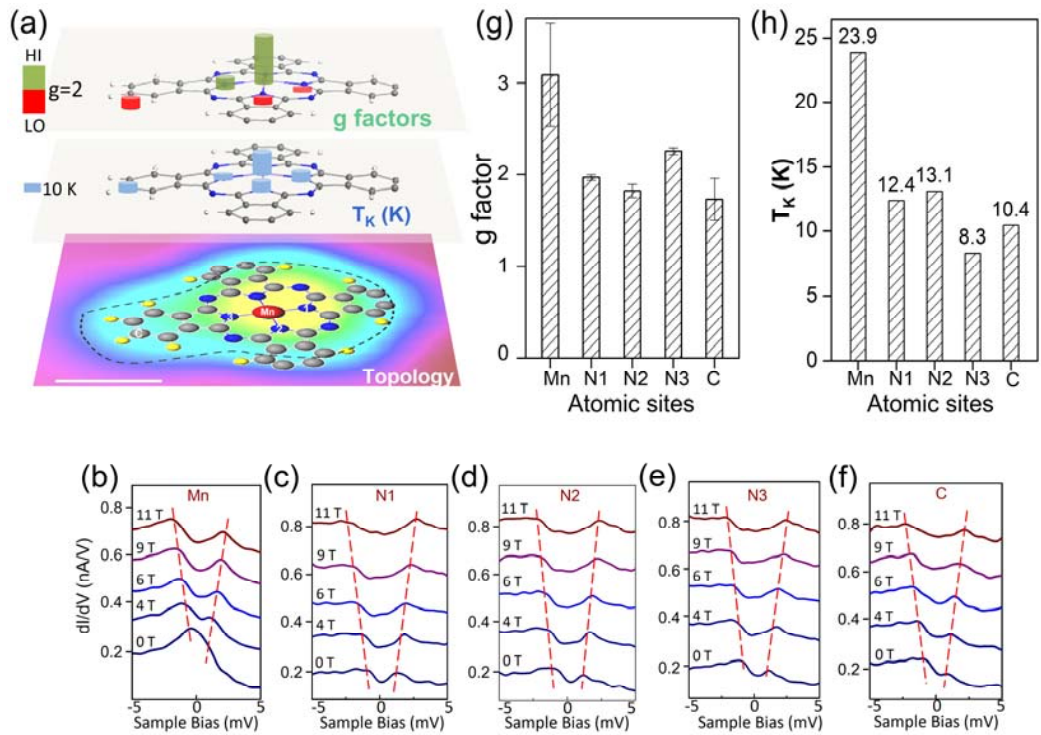

**FIG. 4**